\documentstyle[twoside]{ptptex}
\notypesetlogo  
\title{Scalar Mesons and Chiral States}

\author{%
Muneyuki Ishida$^1$ and Shin {\sc Ishida}$^2$\footnote{Associate member}
}
\inst{%
$^{1}$ Department of Physics, Meisei University, Tokyo 191-8506. Japan\\
$^{2}$ Research Institute of Quantum Science, 
College of Science and Technology\\
Nihon University, Tokyo 101-0062, Japan
}
\recdate{%
\today
}
\abst{%
The essential points and physical backgrounds of the covariant level-classification scheme,
based on $\tilde U(12)_{SF}\bigotimes O(3,1)_L$, are reviewed: 
This scheme is extended from the non-relativistic $SU(6)_{SF}\bigotimes O(3)_L$ scheme 
by introducing the new $SU(2)$-spin ($\rho$-spin) degree of freedom, 
which is necessary for covariant description of composite hadrons. 
Our scheme predicts the existence of new type of chiral mesons and baryons (Chiralons) 
out of the conventional $SU(6)_{SF}\bigotimes O(3)_L$ scheme.
The $\sigma$ nonet is a typical example of chiralons to be assigned to the $(q\bar q)$ 
relativistic $S$-wave state. The new narrow mesons $D_s(2317)/D_s(2463)$ 
are naturally assigned as the ground-state scalar and axial-vector chiralons
in the $(c\bar s)$ system. 

}

\begin{document}

\maketitle
\setcounter{tocdepth}{4}

\section{Introduction}

{\it (Difficulty of Non-Relativistic Classification Scheme)}\ \ \ \ The non-relativistic (NR)
hadron level-classification scheme with the approximate static-symmetry
\begin{eqnarray}
SU(6)_{SF} & \bigotimes & O(3)_L
\label{eq1}
\end{eqnarray}
had been successful for these 4 decades, but recently the necessity for covariant classification has been
strengthened:\ \ Theoretically the QCD, basic dynamics underlying the hadron physics, has the chiral symmetry,
``maximally" relativistic symmetry.
Phenomenologically the property of $\pi$-meson as a Nambu-Goldstone boson in the case of broken chiral symmetry
has been wellknown. Moreover, the existence of light $\sigma$-meson as chiral partner of $\pi$-meson, $\sigma (600)$,
which has been a controversial problem for many years, seems now to be confirmed.
However, there is no suitable seat prepared for it in the NR scheme.

{\it (Covariant Classification Scheme and Chiral states$/$Chiralons)}\ \ \ \ 
Correspondingly to this situation, a few years ago we have proposed\cite{rf2,rf3} a covariant level-classification scheme
for hadrons, unifying the seemingly contradictory two, non-relativistic and extremely relativistic, viewpoints.
Here the framework is manifestly Lorentz-covariant and the space for the static symmetry is extended from
Eq.~(\ref{eq1}) to that of
\begin{eqnarray}
SU(6)_{SF} & \bigotimes & \begin{array}{|c|} \hline SU(2)_\rho \\   \hline \end{array}\  
{\scriptstyle \bigotimes} \ O(3)_L , 
\label{eq2}
\end{eqnarray}
where a new additional $SU(2)$-space for the $\rho$-spin ($\rho$- and $\sigma$- spin being 2 by 2 Pauli-matrices
corresponding to the conventional decomposition of 4 by 4 Dirac $\gamma$ matrices:
$\gamma \equiv \sigma {\scriptstyle \bigotimes} \rho$) is introduced for covariant 
description of hadron spin-wave function (WF). As a result,
in our scheme the \underline{squared-mass} spectra become globally
$\tilde U(12)$-symmetric\cite{rf2,rf4} and the mass spectra themselves are able to be reconciled with the broken
chiral symmetry.
The spin WF for meson and baryon systems are given by the Bargmann-Wigner (BW) spinors,\cite{rf3}
which are represented, respectively, as the bi-Dirac and tri-Dirac spinors
\begin{eqnarray}
\begin{array}{llclcl}
{\rm mesons}  & W_\alpha^\beta               & = & u_\alpha \bar v^\beta & : &
    \alpha =(\rho_3,\sigma_3),\  \beta =(\bar\rho_3,\bar\sigma_3),\\
{\rm baryons} & W_{\alpha_1\alpha_2\alpha_3} & = & u_{\alpha_1} u_{\alpha_2} u_{\alpha_3} & : &
    \alpha_i =(\rho^{(i)}_3,\sigma^{(i)}_3), \\
\end{array}
\label{eq3}
\end{eqnarray}
where $( \alpha ,\beta ,\alpha_i)$ denotes the suffices of Dirac spinors of 
(quark,anti-quark,$i$-th quark) represented by the eigenvalues of $\rho$-spin and $\sigma$-spin.

In light-quark(LL) meson systems the states with the combination of $(\rho_3,\bar\rho_3)$
\begin{eqnarray}
(+,+)\ ;\ {\rm (Pauli-states)},&\ \ & 
  (+,-),(-,+),(-,-)\ ;\ {\rm (Chiral\ states)},\   \label{eq4}
\end{eqnarray}
are expected to be realized in nature, while in the heavy-light quark (HL) meson systems 
only the states with $(\rho_3,\bar\rho_3)=(+,+)$(:Paulons) and $(+,-)$(:Chiralons)
are expetced, reflecting the physical situation that the HL meson system has 
the non-relativistic $SU(6)_{SF}$ spin symmetry (the relativistic chiral symmetry) 
concerning the constituent Heavy quarks (Light quarks). 
The Pauli-states$/$Paulons 
Eq.~(\ref{eq4}) are also describable in NR scheme, while the chiral states$/$Chiralons
Eq.~(\ref{eq4}) have appeared first in the covariant scheme and are out of the NR description.
Similarly in the light-quark baryon systems, the states with $(\rho_3^{(1)},\rho_3^{(2)},\rho_3^{(3)})=(+,+,+)$(:Paulons) and
 $(+,+,-),(+,-,-)$(:Chiralons) are expected to exist.



\section{Description and Level Structure of Hadrons}

{\it (Wave function and Wave Equation)}\ \ \ \ 
The WF of mesons $\Phi_A{}^B$ and baryons $\Phi_{A_1A_2A_3}$ are described systematically as
\begin{eqnarray}
\begin{array}{ll}
\Phi_A{}^B (x,y) \sim  \psi_{q,A}(x) \bar\psi^{q, B} (y), & 
\Phi_{A_1A_2A_3} \sim  \psi_{q_1,A_1}(x_1) \psi_{q_2,A_2}(x_2) \psi_{q_3,A_3}(x_3),\\ 
\ \ \ \ \ \ A = (\alpha ,a),\  B=(\beta ,b); & \alpha ,\beta =(1\sim 4),\ \  a,b=(u,d,s,c,b),\\
\end{array} 
\label{eq6}
\end{eqnarray}
and are assumed to satisfy the Klein-Gordon(KG) equation of Yukawa-type\cite{rf6}
\begin{eqnarray}
\left[  \left( \partial / \partial X_\mu \right)^2 - {\cal M}^2(r_\mu ,\partial /\partial r_\mu \cdots ; \partial / i\partial X_\mu ) \right] 
 \Phi (X,r\cdots )   &=& 0\ .   \label{eq7}
\end{eqnarray}
The operator ${\cal M}^2$ is assumed to contain no light-quark Dirac matrices $\gamma^{(q)}$
in the ideal limit, leading to the chiral symmetric global structure of 
\underline{squared-mass} spectra.
The WF is separated into the two parts, the one of plane-wave center of mass motion and 
the other of internal WF, as
\begin{eqnarray}
\Phi (X,r\cdots ) &=& \sum_N \sum_{ {\bf P}_N, P_{N,0}>0}
\left[  e^{iP_N\cdot X} \psi_N^{(+)} (P_N,r\cdots ) 
     +  e^{-iP_N\cdot X} \psi_N^{(-)} (P_N,r\cdots) \right].\ \ \ \  
   \label{eq8}
\end{eqnarray}

{\it (Expansion of WF on $[$Spinor${\scriptstyle \bigotimes}$Space-time$]$ eigen-functions)}\ \ \ \ 
The internal WF with definite total spin $J$ is expanded in terms of respective eigen functions 
on spinor-space and on internal space-time as
\begin{eqnarray}
\psi_{J,\alpha\cdots}{}^{\beta\cdots }(P_N,r\cdots ) &=& \sum_{i,j} 
c_{ij}^J W_{\alpha\cdots}^{(i)\beta\cdots} (P_N) O^{(j)}(P_N,r\cdots),
   \label{eq9}
\end{eqnarray}
where $W_{\alpha\cdots}^{(i)\beta\cdots} (P_N)$ and $O^{(j)}(P_N,r\cdots )$ are covariant tensors 
respectively, in the $\tilde U(4)_{D.S.}$ (pseudo-unitary Dirac spinor) space and 
the $O(3,1)_L$ (Lorentz-space).

{\it $Spin$ $WF/BW$-$spinors$}\ \ \ \ 
As the complete set of spinor-space eigen-functions we choose the BW spinors, defined as 
solutions of the (local) KG equations
\begin{eqnarray}
\left[ \left( \partial / \partial X_\mu \right)^2 -M^2 \right] W_{\alpha\cdots}^{\beta\cdots} (X) = 0
 & \stackrel{\rm Moment.\ Repr.}{\longrightarrow} & \left( P_\mu^2 + M^2 \right) 
W_{\alpha\cdots}^{(\pm )\beta\cdots} (P) = 0\ \ \ \ \ . 
   \label{eq10}
\end{eqnarray}
For the LL-mesons we have the four physical solutions:\cite{rf3}
\begin{eqnarray}
U_\alpha{}^\beta (P) &\equiv& u_\alpha^{(q)} (P) \bar v_{(\bar q)}^\beta (P);\ 
C_\alpha{}^\beta (P) \equiv u_\alpha^{(q)} (P) \bar v_{(\bar q)}^\beta (-P),\    \label{eq11}\\
D_\alpha{}^\beta (P) &\equiv& u_\alpha^{(q)} (-P) \bar v_{(\bar q)}^\beta (P);\ 
V_\alpha{}^\beta (P) \equiv u_\alpha^{(q)} (-P) \bar v_{(\bar q)}^\beta (-P),\    \label{eq12}
\end{eqnarray}
where $U$ describes the Paulons with $(\rho_3,\bar \rho_3)=(+,+)$, while $C$, $D$ and $V$ describe the
Chiralons with $(\rho_3,\bar \rho_3)=(+,-)$, $(-,+)$ and $(-,-)$, respectively. The $U,V$ have $J^P=0^-,1^-$, while
 $C,D$ have $J^P=0^+,1^+$. 
The eigen-functions of charge conjugation parity are obtained through the superpositions, 
$W=U\pm V$ and $C\pm D$. Their explicit forms are given by
\begin{eqnarray}
{\scriptstyle 
\begin{array}{rcccccccc}
W_\alpha{}^\beta (v): & i\gamma_5 & i\tilde\gamma_\mu & -\gamma_5v\cdot\gamma & -i\sigma_{\mu\nu}v_\nu 
                              & 1 & i\gamma_5\tilde\gamma_\mu & -v\cdot\gamma & -\gamma_5\sigma_{\mu\nu}v_\nu \\   
               J^{PC}: &  0^{-+}  &  1^{--}  &  0^{-+} &  1^{--}
                             &  0^{++}  &  1^{++} &  0^{+-}  &  1^{+-} \\    
             \phi :  &  P_s^{(N)}  &  V_\mu^{(N)}  &  P_s^{(E)} &  V_\mu^{(E)}
                             &  S^{(N)}  &  A_\mu^{(N)} &  S^{(E)}  &  A_\mu^{(E)} \\
\end{array}  }  \label{eq13}
\end{eqnarray}  
The above $W_\alpha{}^\beta (v)$ are the tensors of $\tilde U(4)(\supset SU(2)_\rho \bigotimes SU(2)_\sigma )$ 
symmetry, and correspond to all the 16 Dirac $\gamma$-matrices. 
By including the flavor SU(3), the LL mesons are classified as 
$\underline{144(=12\times  12^* )}$ representation of 
$\tilde U(12)_{SF}(\supset  SU(3)_F\bigotimes SU(2)_\rho \bigotimes SU(2)_\sigma )$.\cite{rf2,rf4}
In terms of static $SU(6)_{SF} (\supset  SU(3)_F \bigotimes SU(2)_\sigma  )$, the \underline{144} includes 
four $\underline{36(=6\times 6^*)}$. 

As is evidently seen from Eq.~(\ref{eq13})
and from the chiral transformation on light quarks reducing 
$\gamma_5 u(P) = u (-P)$ and $\bar v(P)\gamma_5=\bar v (-P)$, 
the $(P_s^{(R)},S^{R})$ and $(V_\mu^{(R)},A_\mu^{(R)})$ $(R=N,E)$ 
form linear representations of chiral symmetry. 

For the HL-mesons we have the two physical solutions, $U$ and $C$, given in Eq.~(\ref{eq11}). 
They are decomposed into the pseudo-scalars$/$vectors, and scalars$/$axial-vectors, respectively, as
\begin{eqnarray}
U_\alpha{}^\beta  &\sim& (1-iv\cdot\gamma )
\left[ i\gamma_5 P_s + i\tilde\gamma_\mu V_\mu \right],
\ \ 
C_\alpha{}^\beta \sim (1-iv\cdot\gamma )\left[ S + i\gamma_5\tilde\gamma_\mu A_\mu \right].
\ \ \ \ \ \
\label{eq14}
\end{eqnarray}
The $U$ and $C$ form $\underline{12^*}$ of $\tilde U(12)_{SF}$ symmetry.
Through the chiral transformation, the former is changed into the latter as $U(P)\gamma_5=C(P)$.

The light quark ground state baryons are assigned as $\underline{(12\times  12 \times 12)_{Sym} =364}$ 
representation of $\tilde U(12)_{SF}$ symmetry. The \underline{364} includes the baryons ($\underline{182_B}$) 
and anti-baryons ($\underline{182_{\bar B}}$). 
In terms of static $SU(6)_{SF}$, the $\underline{182_B}$ includes the 
$\underline{56_E}$(:Paulons) and $\underline{70_G}$, $\underline{56_F}$ (:Chiralons) with 
$(\rho_3,\rho_3,\rho_3)=(+,+,+)$ and $(+,+,-),(+,-,-)$, respectively. 
Both $\underline{56_{E}}$ and $\underline{56_{F}}$ include $N$-octets and $\Delta$-decouplets with positive parity,
while the negative parity $\underline{70_G}$ include 
$N$-octets with $J=1/2$ and $3/2$, $\Delta$-decouplet with $J=1/2$ and $\Lambda$-singlet with $J=1/2$.

{\it $Internal$ $space$-$time$ $WF/Yukawa$ $oscillators$}\ \ \ \ 
As the complete set of space-time eigen-functions we choose the covariant, 4-dimensional Yukawa
oscillator functions. By imposing the freezing relative-time condition they become effectively the conventional,
3-dimensional oscillators:
\begin{eqnarray}
\langle P_\mu r_\mu \rangle =  \langle P_\mu  p_\mu \rangle = 0 & \Rightarrow & O(3,1)_L \approx O(3)_L\ .\ \ \   
\label{eq15}
\end{eqnarray}

({\it $Mass$ $spectra$ $for$ $low$-$lying$ $mesons$ $and$ $baryons$)}\ \ \ \ 
Since of the static symmetry (\ref{eq2}) the global mass spectra are given by
\begin{eqnarray}
M_N^2 &=& M_0^2 + N \Omega ,\  N\equiv 2n+L  ,   
\label{eq16}
\end{eqnarray}
leading to phenomenologically well-known Regge trajectories.

The masses of ground state mesons and baryons are degenerate in the ideal limit, 
and they split with each others between chiral partners (spin partners) 
by the bilinear scalar-quark condensates (the perturbative QCD spin-spin interaction).

\section{Candidates for Chiralons$/$Concluding Remarks}

({\it Experimental candidates of chiral particles})\ \ \ \ 
In our level-classification scheme a series of new type of multiplets, 
chiralons, are predicted to exist in the ground and 
the first excited states. Presently we can 
give only a few experimental candidates or indications for them:

\underline{\em (LL-mesons)}\ \ \ \ 
One of the most important candidates is the scalar $\sigma$ nonet to be assigned as $S^{(N)}(^1S_0)$ : 
$[\sigma (600),\ \kappa (900),\ a_0(980),\ f_0(980)]$. The existence of $\sigma (600)$ seems to be 
established\cite{rf1} through the analyses of, especially, $\pi\pi$-production processes. 
The firm experimental evidences for $\kappa (800$-900) were reported in the production 
processes\cite{rf12}\cite{rf13}. The properties of $\kappa$ are consistent with those given formerly in
$K\pi$ scattering phase shift\cite{rf11}.\\
In our scheme the two sets of $P_s$- and of $V_\mu$-nonets are to exist: 
The vector mesons\cite{rf14} 
$\rho (1250)$ and $\omega (1250)$, suspected to exist for long time, are naturally able to 
be assigned as the members of $V_\mu^{(E)}(^3S_1)$-nonet.\\
Out of the three established $\eta$, $[\eta (1295),\ \eta (1420),\  
\eta (1460) ]$ at least one extra, plausibly $\eta (1295)$ with the 
lowest mass, may belong to $P_s^{(E)}(^1S_0)$ nonet.\\
The two ``exotic" particles $\pi_1(1400)$ and 
$\pi_1(1600)$ with $J^{PC}=1^{-+}$ and $I=1$, observed\cite{rf15}   
in the $\pi\eta ,\  \rho\pi$ and other channels, 
may be naturally assigned as the first excited states $S^{(E)}(^1P_1)$ and $A_\mu^{(E)}(^3P_1)$ 
of the chiralons.

(\underline{HL-mesons})\ \ \ \ 
The new narrow mesons $D_s(2317)/D_s(2463)$, observed recently in the final states 
$D_s^+ \pi^0/D_s^{*+} \pi^0$, are pointed out\cite{rf16} to be naturally assigned as the ground-state 
scalar and axial-vector chiralons in the $(c\bar s)$ system.
Their decay properties are well-explained in our scheme.
The scalar$/$axial-vector chiralons are also expected to exist in $c\bar n$ system, denoted as  
$D_{n0}^\chi /D_{n1}^\chi $. Their masses are predicted around $2110/2250$ MeV
by using SU(3) linear $\sigma$ model. 
Some results in search\cite{rf17} for their existence are reported.

(\underline{\em $qqq$}-baryons)\ \ \ \  The two facts have been a 
longstanding problem that the Roper resonance $N(1440)_{1/2^+}$ 
is too light to be assigned as radial excitation of $N(939)$ and 
that $\Lambda (1405)_{1/2^-}$ is too light as the $L=1$ excited 
state of $\Lambda (1116)$. In our new scheme these two problems dissolve\cite{rf18} in principle,
because in the ground states there exist the two \underline{56} with positive parity,
$\underline{56_E}^{+\hspace{-0.3cm}\bigcirc}$ and $\underline{56_F}^{+\hspace{-0.3cm}\bigcirc}$,
and one negative-parity $\underline{70_G}$.
The other puzzle, that the predicted width 
$\Gamma (\Delta\rightarrow N\gamma)$ in the conventional treatment is much
small compared with the experiment, may also be solved by considering 
the relativistic effect of the mixing between $\underline{56_E}$ and $\underline{56_F}$.
The other problem of 
extremely small width of $\Delta (1600) \rightarrow N\gamma$ is explained
by the orthogonality between WFs of $\Delta (1600)$ and of $\Delta (1232)$
by assuming them to be ground states.\cite{rf18}

({\it Concluding remarks})\ \ \ \ 
We have summarized in this talk the essential points of the covariant level-classification scheme,
which has, we believe, a possibility to solve the serious problem in hadron spectroscopy mentioned 
in Introduction.
In this connection further investigations, both experimental and theoretical, for chiral states 
predicted in this scheme, are urgently required for new development of hadron physics.


\begin{thebibliography}{99}

\bibitem{rf2} S.~Ishida and M.~Ishida, Phys. Lett. B{\bf 539} (2002) 249.
\bibitem{rf3} S. Ishida, M. Ishida and T. Maeda, Prog.~Theor.~Phys. {\bf 104} (2000), 785.
\bibitem{rf4} A. Salam, R. Delbourgo and J. Strathdee, Proc. R. Soc. London {\bf A284} (1965) 146.\\
             B.~Sakita and K.~C.~Wali, Phys.~Rev.~{\bf B139} (1965) 1335.
\bibitem{rf6} H. Yukawa, Phys. Rev. {\bf 91} (1953), 415, 416. 



\bibitem{rf1} N.~A.~T\"ornqvist, summary talk in proceedings of ``$\sigma$-Meson 2000",
KEK-proceedings 2000-4; NUP-B-2000-1; Soryusiron kenkyu {\bf 102} (2001) No.5.
\bibitem{rf11} S. Ishida et al., Prog.~Theor.~Phys. {\bf 98} (1997), 621.\\ 
                E. Beveren et al., Z. Phys. {\bf C30} (1986), 651.
\bibitem{rf12} T.Komda in this proceedings. Wu Ning, hep-ex/0304001. 
\bibitem{rf13} C.~Gobel, in proc. of Nihon univ. and KEK symp., Ichigaya, Tokyo, Feb 24-26, 2003.
\bibitem{rf14} M.~Oda, in this proceedings.
\bibitem{rf15} S. U. Chung, summary talk of Hadron99.\  
               V. Dorofeev; A. Popov, proc. of Hadron2001.
\bibitem{rf16} S. Ishida, in this proceedings.\ 
               M. Ishida and S. Ishida, Prog. Theor. Phys. {\bf 106} (2001), 373.
\bibitem{rf17} I.~Yamauchi, in this proceedings.
\bibitem{rf18} M.~Ishida, in proc. of Nihon univ. and KEK symposium, 2003.

\end{thebibliography}
\end{document}